# Temperature Dependent Failure of Atomically Thin $MoTe_2$

*A S M Redwan Haider[1#], Ahmad Fatehi Ali Mohammed Hezam[1#], Md Akibul Islam[2*], Yeasir Arafat[3], Mohammad Tanvirul Ferdaous[3], Sayedus Salehin[1], Md.Rezwanul Karim[1*]*

[1]Depatment of Mechanical and Production Engineering, Islamic University of Technology, Bangladesh.

[2]Department of Mechanical and Industrial Engineering, University of Toronto, Canada.

[3]Independent Researcher.

[#]Equal Contributions

[*]Corresponding Authors

## Abstract:

In this study, we systematically investigated the mechanical responses of monolayer molybdenum ditelluride ($MoTe_2$) using molecular dynamics (MD) simulations. The tensile behavior of trigonal prismatic phase (2H phase) $MoTe_2$ under uniaxial strain was simulated in the armchair and zigzag directions. We also investigated the crack formation and propagation in both armchair and zigzag directions at 10K and 300K to understand the fracture behavior of monolayer $MoTe_2$ at different temperatures. The MD simulations show clean cleavage for the armchair direction and the cracks were numerous and scattered in the case of the zigzag direction. Finally, we investigated the effect of temperature on Young's modulus and fracture stress of monolayer $MoTe_2$. The results show that at a strain rate of $10^{-4}$ $ps^{-1}$, the fracture strength of monolayer $MoTe_2$ in the armchair and zigzag directions at 10K is 16.33 GPa (11.43 N/m) and 13.71 GPa (9.46 N/m) under a 24% and 18% fracture strain, respectively. The fracture strength of monolayer $MoTe_2$ in the armchair and zigzag direction at 600K is 10.81 GPa (7.56 N/m) and 10.13 GPa (7.09 N/m) under a 12.5% and 12.47% fracture strain, respectively.

**Graphical Abstract:**

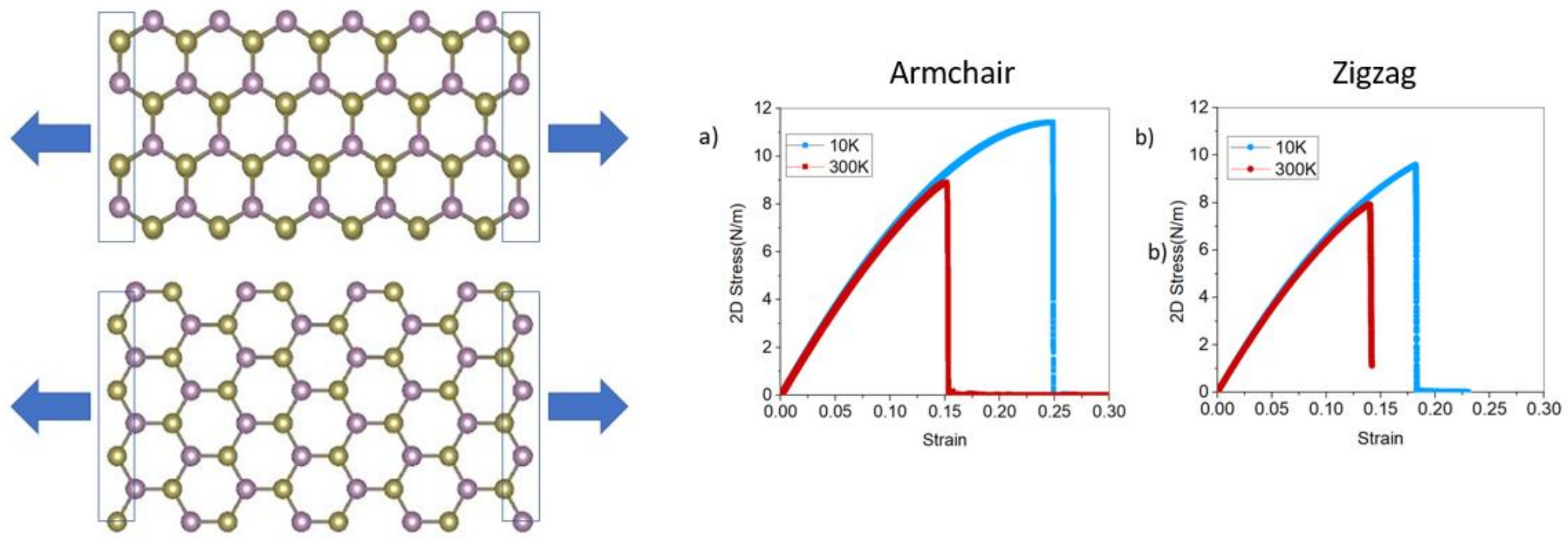

**Keywords**: 2D materials, Molybdenum ditelluride ($MoTe_2$), Molecular Dynamics (MD), Stillinger-Weber (SW) potential, Fracture mechanics.

…………………………………………………………………………………………………

## 1. Introduction

Since the discovery of graphene in 2004[1], atomically thin two-dimensional materials have garnered great interest due to their excellent mechanical [2], electrical [3], thermal [4], optical [5], anti-corrosive [6], and magnetic properties [7]. While graphene has been demonstrated as the strongest material ever measured, its inherent lack of a band gap has constrained its utility in the semiconductor industry, thus it is difficult to use graphene in transistor applications [1] and bio-electronics [8]. Transition metal dichalcogenides (TMDCs) hence emerged as a promising candidate for transistor applications due to the presence of direct and indirect band gaps in their atomic structure [9], [10], [11]. A TMDC is a sandwich structure where a metal atom (M) is sandwiched between two chalcogen atoms (X) and is represented as $MX_2$ ($MoS_2$, $MoTe_2$, $WS_2$ etc.). Group VI TMDCs (Mo and W) are of particular interest due to more than one possible polymorph. These materials have become a key component in the flexible and wearable electronics industry [12], [13] . TMDCs have specific mechanical and structural properties that enable researchers to develop lateral hetero-structures which give rise to flexibility in the development of electronic devices [14], [15], [16]. The preliminary studies conducted on the mechanical characteristics of TMDCs were focused on obtaining properties such as Young's modulus of pristine $MoS_2$ obtained using exfoliation techniques [17], [18], [19]. One of these structural phases exhibits a stable semiconducting phase (2H), while the other two demonstrate metallic phases with monoclinic (1T′) and orthorhombic ($T_d$) structures[20], [21], [22]. This is an interesting property since the semiconducting phase can be used to make electronic devices [20] and the metallic phase can be used in catalytic activities leading to hydrogen evolution [23], [24], [25]. It has been suggested in the literature that the phase transition between the semiconducting and metallic phases is possible using controlled strain [26]. Despite attracting the most interest from the research community, $MoS_2$ may not be the best candidate for strain-induced phase transition due to the high energy required for the transition ($\Delta E > 0.8$ eV) [26]. In contrast, $MoTe_2$ exhibits a significantly smaller energy difference between its semiconducting and metallic phase ($\Delta E < 50$ meV) [26], [27]. Hence, it is of utmost importance to investigate the mechanical properties of monolayer $MoTe_2$ under controlled strain to determine the stability and durability of this TMDC. Based on optical absorption spectra at a temperature of 4.2 K, it has been determined that 2H-$MoTe_2$ possesses an indirect band gap of approximately $1.06 \pm 0.01$ eV. In contrast, 1T'- $MoTe_2$ displays semi-metallic behavior under the same conditions [28]. $MoTe_2$ has recently been used in a wide range of applications ranging from photo-detectors [29], and energy storage [30] to novel piezo-electric devices [31]. Claudia et al. [32] have characterized the single layer and multilayer $MoTe_2$ by using photo-luminescence, optical absorbance, and Raman scattering. The results suggest that the monolayer of $MoTe_2$ has an optical band gap of 1.10 eV which is quite high in comparison to bulk or multi-layered $MoTe_2$. The use of $MoTe_2$ for the advancement of photodetectors and solar cells presented fascinating findings, such as the achievement of higher power efficiency to up to 8.94% due to the addition of 10 vol% proportion of $MoTe_2$ nanostructures in PBDB-T: ITIC polymer matrix as an active layer [28]. TMDCs have been incorporated in field-effect transistors since 2004, as they exhibited ambipolar behavior with high mobility and current modulation. Conan et al. [33] deduced that the mobility of the bulk $MoTe_2$ can reach up to 200 $cm^2\ V^{-1}\ s^{-1}$ at room temperature compared to the mobility of $MoS_2$ at room temperature which is 100 $cm^2\ V^{-1}\ s^{-1}$ [34]. Controlled growth conditions were used to synthesize materials on $MoTe_2$ crystals for the development of light-emitting transistors [35]. Another fascinating property of monolayer $MoTe_2$ is its flexibility and stretchability, making it possible to apply external deformation to manipulate their physical properties [29]. The flexibility allows the device to be subjected to bending and stretching and thus can be used to manufacture bendable screens, electronic paper, electronic skin, and wearable bio-sensors [36], [37]. Among all the TMDCs, $MoTe_2$ has attracted special attention due to the possibility of phase transition between two stable phases- semiconducting 2H (hexagonal; ~1 eV bandgap) and metallic 1T (octahedral) [27], [38], [39], [40]. The first principal energy calculations revealed that $MoTe_2$ has a smaller energy difference between 2H and 1T phases than other TMDCs. The literature has reported that monolayer 2H-$MoTe_2$ can be phase transitioned to 1T- $MoTe_2$ using only 3% strain [26]. This paves the way for the low energy transition between semiconducting and metallic phase which can be useful in-memory applications [41],[41]. Atomic force microscopy (AFM) has been recently used to demonstrate the phase transition in $MoTe_2$ using the nanoindentation method [42]. However, due to the instability of $MoTe_2$ in ambient air [43], it is difficult to conduct a detailed study on the phase transition and the mechanics of monolayer $MoTe_2$. MD simulation offers a unique opportunity to perform experiments on the 2D $MoTe_2$ in a controlled environment. Multiple simulations can be conducted on the same material to predict the outcome before performing hands-on experiments. However, it is worth mentioning that atomic force microscopy (AFM) based nanoindentation techniques have been used to measure the elastic properties of a few layers of $MoTe_2$ [44] and not the monolayers which are of most interest. The mechanical response of 2H, 1T, and 1T'-$MoTe_2$ to uniaxial tensile conditions was investigated using DFT [45]. Initially, these 2D layers demonstrated a linear response, which was followed by a non-linear trend up to the ultimate tensile stress. Understanding the fracture mechanics of TMDCs such as $MoTe_2$ is essential for the design and engineering of nano-devices [46], [47], [48], [49]. The fracture patterns due to different stress conditions help to predict the reliability and durability of such devices. Predicting the failure modes will help researchers prevent catastrophic failures by identifying the weak points and developing preventive measures [49]. The purpose of this work is to discuss the deformation and fracture behavior of monolayer $MoTe_2$ at a wide range of temperatures. The stress-strain response of 2D materials is unique compared to their bulk counterpart [50]. To mimic the mechanical properties of $MoTe_2$ the parameterized SW interatomic potential developed by Jiang et al. [51] was used. Then we applied uniaxial strain [52] using molecular dynamic (MD) simulation along the zigzag and armchair direction until the monolayer failed. Analysis of stress-strain responses beyond the linear elastic regime reveals directional anisotropy in mechanical behavior and failure qualities. Finally, we varied the temperature from 100K to 600K to investigate the elastic and failure properties in monolayer $MoTe_2$. The computational findings presented in this work are consistent with existing literature in terms of the elastic properties of $MoTe_2$ [51], [53]For monolayer $MoTe_2$ in both armchair and zigzag directions, the fracture stresses, fracture strengths and Young's modulus decrease as the temperature increases.

One of the stable phases of monolayer $MoTe_2$ is the 2H phase, as shown in Figure 1. Here hexagonally arranged, each molybdenum atom is sandwiched between two layers of tellurium atoms forming a Te-Mo-Te sequence. Each Mo atom has six Te atoms as its first neighbors, and each Te atom is directly bound to three Mo atoms in the trigonal prismatic coordination of the intralayer atoms (Mo and Te atoms within a single layer). Te and Mo are bonded by strong covalent bonds. The covalent bonds are predominantly governed by three bond-stretching (Mo–Te, Mo–Mo, and Te–Te) and three angle-bending motions (two Mo–Te–Te angles with Mo being the central atom and one Te–Mo–Mo). For the convenience of creating the structures in MD, we used a hexagonal unit cell (a = b = 3.159 Å and c=13.964 Å, α = β = 90° and γ=120° [54]).

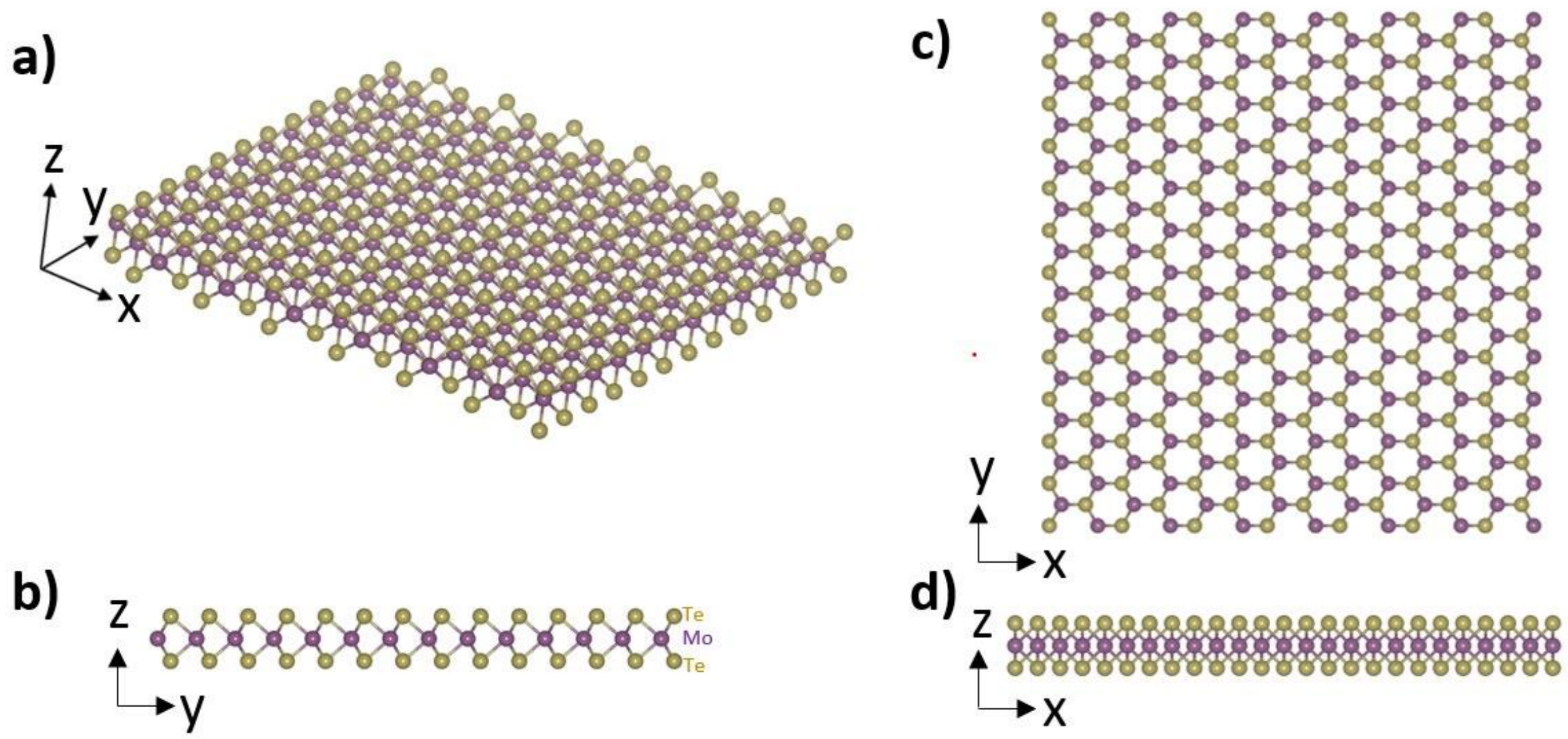

Figure 1: The crystal structure of monolayer $MoTe_2$ showing a layer of molybdenum atoms (purple) sandwiched between two layers of Tellurium atoms (olive). a) Orthogonal View, b) Front View, c) Top View and d) Side View.

## 2. Computational Method

In this section, we introduce the simulator software that was used to perform the simulations along with the essential working conditions and environment that were required to run the simulation. The description of the $MoTe_2$ structure that was used is also provided describing the bond lengths, angles, and total dimensions of the sheet. Additionally, the governing equations of the model are provided.

### 2.1 Modelling monolayer $MoTe_2$

*LAMMPS (Large-scale Atomic/Molecular Massively Parallel Simulator*) was used to carry out MD simulations [55]. It can model nano-scale to macroscopic systems using various atomic potentials and boundary conditions. In our work, Stillinger–Weber (SW) potential is used [51], which describes the interaction of atoms inside a TMDC and the coupling between them. SW potential parameters are parameterized for different materials [45] to develop a better interatomic potential, for describing the structural, mechanical and phonon spectrum of several materials [48],[55]. There are two inequivalent bond angles in any hexagonal molybdenum dichalcogenides unit cell. Since the values are very close, a single bond angle is considered for simplification of the calculation. The modification done in the literature gives a reasonably accurate result for $MoS_2$ [56]. After the modification, Jiang et. al [56] tested the strain-induced buckling on a composite TMDC heterostructure. The parameterized SW potential was also used to perform simulation on materials such as monolayer Indium Selenide (InSe) [57], Tungsten Di selenide ($WSe_2$) [58] and Borophene [59].

Figure 1 shows that the 2H hexagonal $MoTe_2$ structure has a Molybdenum (Mo) connected to a trigonal prism of six tellurium. The bond length of Mo-Te is approximately taken as 2.73 Å. Each Tellurium (Te) is surrounded by ten other Te atoms. Six Te atoms have a bond length of 3.52 Å in the plane (001), one directly below is at 3.63 Å and three above it is at 3.92 Å [54]. The bond angles of Te-Mo-Te are 83.5°, 80.4° and 133.9° [54]. However, due to the parameterization of SW potential, the two inequivalent bond angles are taken as equal to 80.581° [60]. In the simulation, a 30 nm x 30 nm $MoTe_2$ sheet is used which was prepared using VESTA [61] that contains 25800 atoms in total [51] The thickness of the monolayer of $MoTe_2$ is considered to be 0.7 nm [62]. Periodic boundary conditions are applied in all directions, and a vacuum thickness of 30 Å is used to prevent interactions of atoms at opposite edges.

In our computational study, we utilized the LAMMPS simulator to conduct molecular dynamics simulations on a monolayer $MoTe_2$ sheet. We defined several variables essential to the simulation, such as temperature, which we varied from 10 K to 600 K. Other parameters such as pressure (1 bar), timestep (set to 0.001 fs), strain rate ($10^{-4}$ $ps^{-1}$) were kept constant for each simulation of each separate temperature [63], [64].

In our approach, we employed a two-stage optimization process. First, we performed a geometry optimization at 0 K temperature using the conjugate gradient method. We then thermally equilibrated the system in an NPT ensemble for various time durations as required for a stable material. After relaxation, we applied a uniform uniaxial tensile force to the $MoTe_2$, inducing a uniform displacement across the structure. During the MD simulation, we utilized an NPT ensemble, maintaining the number of atoms, system pressure, and temperature constant, while the volume was allowed to fluctuate. The damping constants for both the temperature and pressure are 20 picoseconds. The relaxation of the structure for temperature, potential energy, and total energy with each step refers to how these properties evolve as the system reaches a state of equilibrium. The strain is applied in the y-direction while the box boundaries fluctuate in the x-direction, preserving the desired temperature and pressure. During this phase, the system's response to the applied strain, such as changes in the pressure tensor and the strain value itself, was continuously monitored and recorded.

## 2.2 Stress-strain relation and distribution

The stress-strain curves were obtained by deforming the simulation box uniaxial and calculating the average stress over the structure. Atomic stress was calculated based on the definition of virial stress. Virial stress components are calculated using [65], [66]:

$$\sigma_{ij}^{Virial} = \frac{1}{\Omega} \sum_{\mathrm{i}} [(-m_i \dot{u}_i \otimes \dot{u}_i + \frac{1}{2} \sum_{j \neq \mathrm{i}} r_{\mathrm{i}j} \otimes f_{ij})] \qquad (1)$$

The summation is upon all the atoms in the total volume $\Omega$ , $m_i$ is the atomic mass $i$, $\dot{u}_i$ is the time derivative which indicates the displacement of the atom concerning a reference position, $r_{ij}$ is the position vector of the atom, $\otimes$ is the cross product, and $f_{ij}$ is the interatomic force applied on atom $i$ by atom $j$. Here, the Stillinger-Weber (SW), potential was used to define the interatomic interactions [51]. The SW potential comprises two-body and three-body terms describing bond stretching and breaking.

The mathematical expressions used to calculate the components of the virial stress:

$$\phi = \sum_{i<j} V_2 + \sum_{i>j<k} V_3 \quad (2)$$

$$V_2 = Ae^{\left[\frac{\rho}{r-r_{max}}\right]}\left(\frac{B}{r^4} - 1\right) \quad (3)$$

$$V_3 = K\varepsilon e^{\frac{\rho_1}{r_{ij}-r_{maxij}} - \frac{\rho_2}{r_{ik}-r_{maxik}}} (cos\, cos\, \theta \, - cos\, cos\, \theta_0)^2 \quad (4)$$

Here $V_2$, and $V_3$ are the two-body bond stretching and angle bending terms accordingly. The terms $r_{max}$, $r_{maxij}$, $r_{maxik}$ are cut-off ratios and $\theta_0$ at the equilibrium configuration is the in-between angle of the two bonds. A and K are energy-related parameters based on the Valence Force Field (VFF) model. $B, \rho, \rho_1$ and $\rho_2$ are other parameters that are fitted coefficients. A and K are two energy parameters. Such parameters along with the corresponding values, were extracted from Li et. al [51], [67].

To calculate the plane stress, we used equation 5. The stress calculated in equation 1 was divided by the thickness of $MoTe_2$ (0.7 nm) to calculate the plane stress using:

$$Plane\ Stress = \frac{Stress}{thickness} \quad (5)$$

## 3. Results and Discussion

To investigate the fracture process in the monolayer $MoTe_2$, we subjected the sheet under uniaxial tension in the armchair and zigzag directions (as shown in Figure 27) and increased the strain until the sheet was completely fractured.

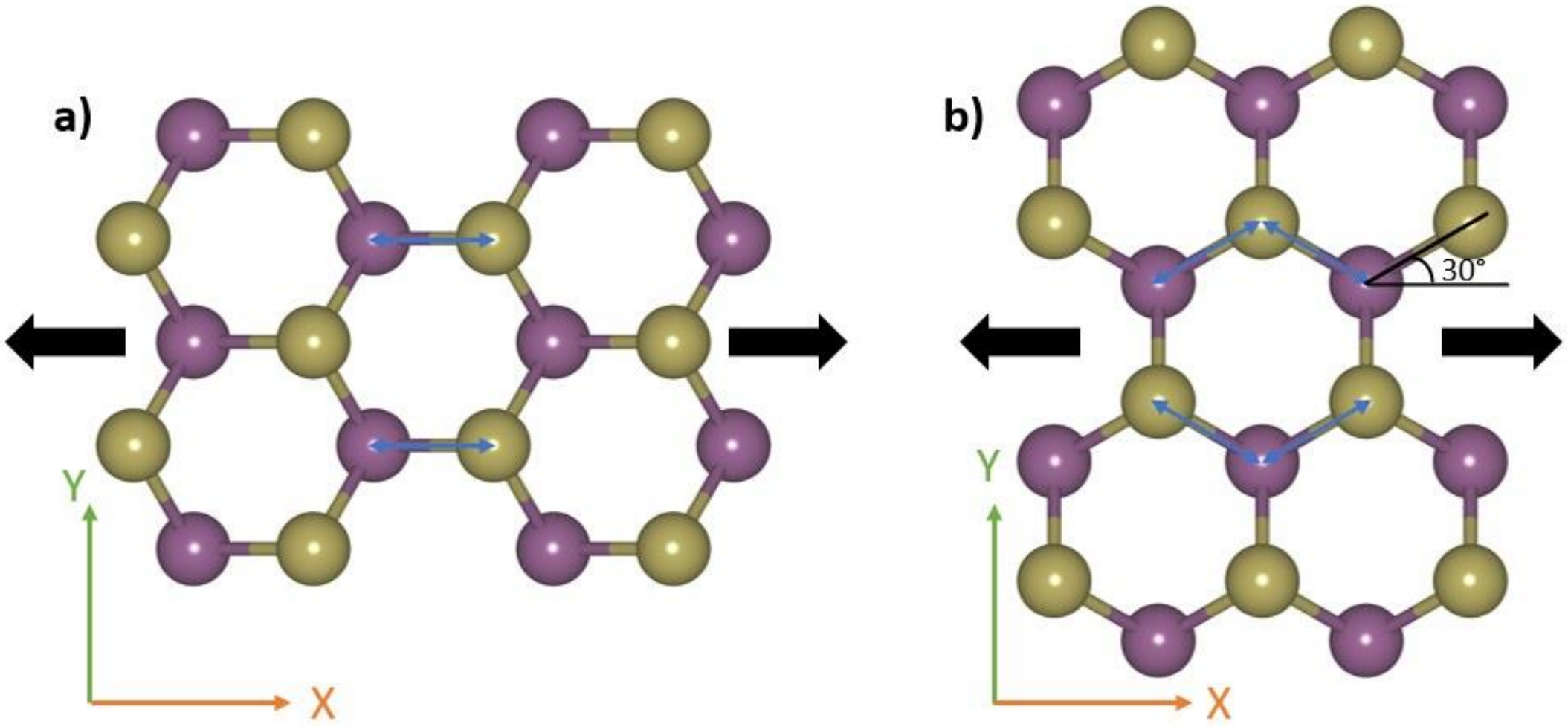

Figure 2: Planar view of the bonds in the structure that are affected most by the applied stress of $MoTe_2$ when the loading is in (a) armchair direction and (b) zigzag direction. The Blue arrows show the affected bonds during propagation.

We used 10K and 300K temperatures initially to understand the effect of temperature on the fracture behavior of the 2D $MoTe_2$ sheet. At 10K, we strained the 2D $MoTe_2$ along the armchair and zigzag edges from 0% to 24.83% at a strain rate of $10^{-4}$ $ps^{-1}$ and the stress in the 2D $MoTe_2$ increased from 0 to ~12 GPa. Although the entire sample surface is subjected to uniform tension deformation, thermal variations at the low temperature cause the atomic stresses not to be distributed uniformly. Using the visualizations created by OVITO [68], it was evident that the crack propagation varied significantly between armchair and zigzag directional loading.

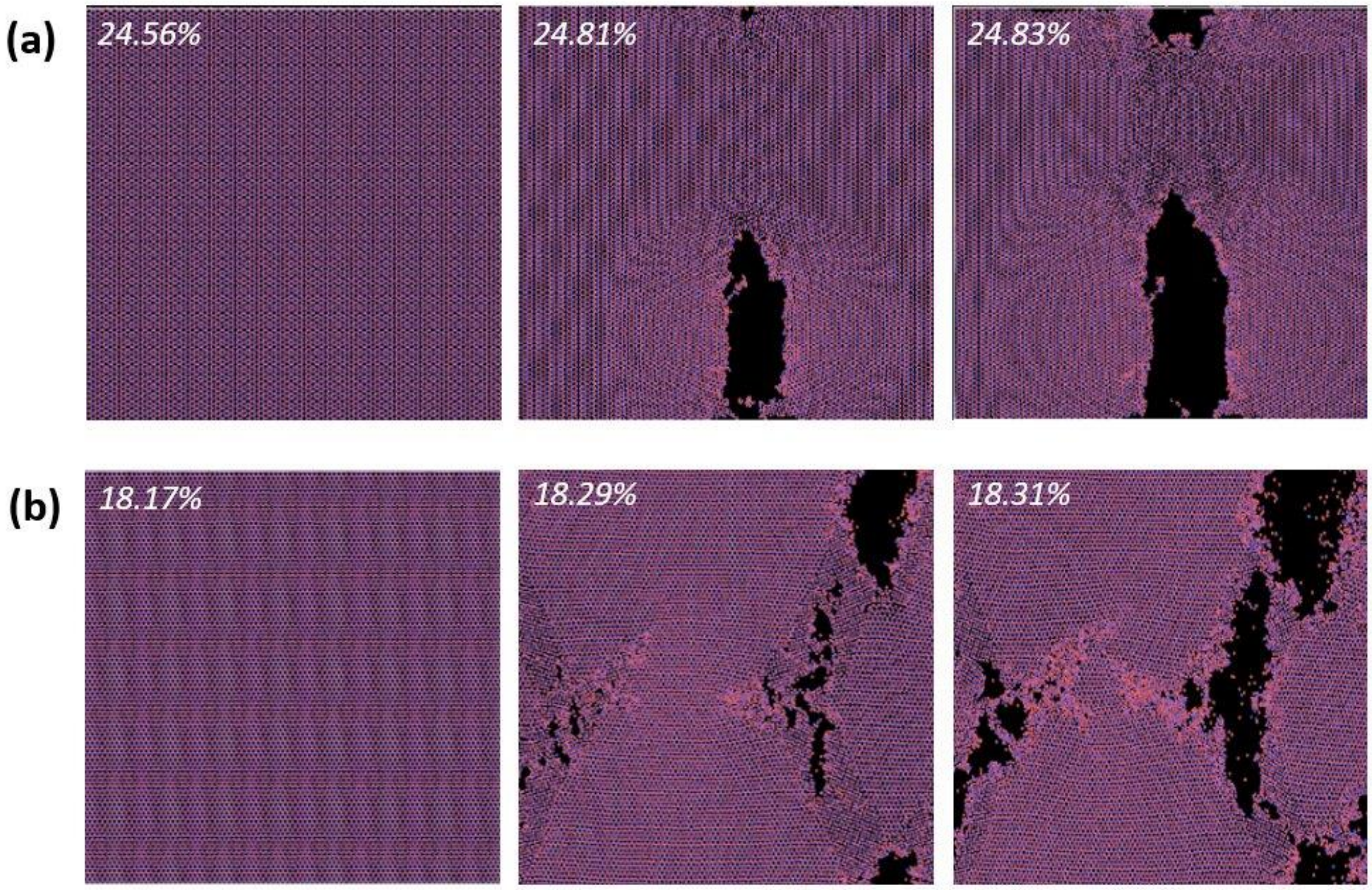

Figure 3: Crack nucleation and propagation in 2D $MoTe_2$ along the (a) armchair and (b) zigzag edges at 10K.

It was observed in Figure 3 (a) that along the armchair direction, a small crack starts to nucleate at 24.81% strain by breaking atomic bonds. The crack starts to propagate as the strain is increased without showing any sign of plasticity in the 2D $MoTe_2$ sheet. Most pristine 2D materials tend to exhibit brittle failure when subjected to additional strain [69]. Up to 24.83% strain, although a crack had already been initiated, it did not yet propagate enough through the material to cause a complete fracture. A similar characteristic was observed along the zigzag edges at 10K temperature Figure 3(b). The applied strain was increased from 0% to 18.31% and the stress in the 2D sheet increased from 0 to ~10 GPa. A small crack started to nucleate at 18.17% strain which caused a brittle failure, and the sheet completely failed at 18.31% strain. It is to be noted that, unlike the armchair edge, the fracture along the zigzag edge was not a clean cleavage. This demonstrated that the fracture occurred preferentially along a puckered groove for loads in the armchair direction rather than across the grooves. Zigzag edge images suggested that the material tends to fracture diagonally across the sheet of material due to the distribution of force components in two directions as demonstrated in the next section. A similar fracture pattern was observed in the case of uniaxial stretching of single-layer $MoS_2$ (zigzag orientation) [70].

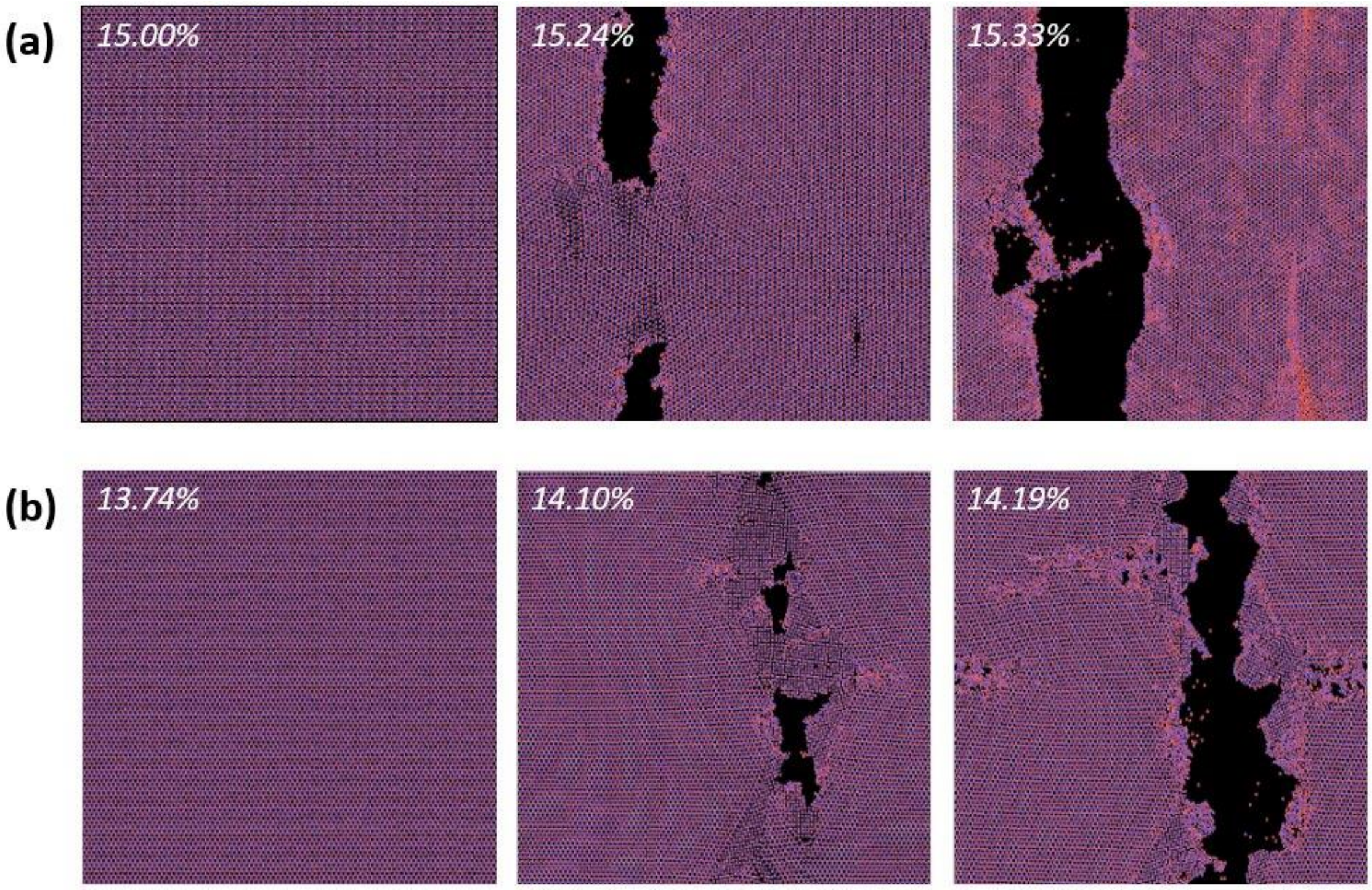

Figure 4: Crack nucleation and propagation in 2D $MoTe_2$ along the (a) armchair and (b) zigzag edges at 300K

Afterwards, we increased the temperature to 300K and applied a similar uniaxial strain on the 2D $MoTe_2$ sheet along the armchair and zigzag edge as shown in Figure 4. For the armchair direction, the crack started to nucleate at 15.2% strain and then cleaved suddenly into two clean pieces at 15.33% strain (Figure 4a). The 2D fracture stress recorded was ~8.5 N/m. However, in the zigzag direction, the crack started to nucleate at 14.1% strain and failed at 14.19%. The fracture 2D stress was recorded at ~8 N/m (Figure 4b). In both Figures 3 and 4 rippling was observed due to the wrinkling of outer Te layers. Since the poison's ratio is positive, when the 2D sheet is stretched on one side it produces a compression in the other direction [71]. Therefore, the compression leads to a wrinkling effect.

We observed a difference in the fracture mechanism of $MoTe_2$ due to the armchair and zigzag loading conditions. It was observed that the armchair and zigzag loading have a perceptible difference in crack initiation and propagation as uniaxial tensile strain is applied.

The images in Figure 2 portrayed the loading conditions on both armchair and zigzag orientations and the force effect on the bonds to initiate and propagate cracking. The results suggest that the crack initiated at a relatively higher strain in the case of armchair loading in comparison to zigzag loading. We have analyzed the bonds and the angles between Mo-Te from a planar view. The bond length between Mo and Te atoms from a 2D perspective is parallel to the loading direction in the case of armchair loading. Thus, the bond lengths stretch at a 0° angle creating a clean cleavage while fracturing. However, the zigzag loading is at 30° with the direction of loading therefore the applied stress becomes a part of the actual stress in the direction of loading. Hence, there is a diagonal fracture pattern in the sheet of the material.

Furthermore, the images in Figure 3 suggest that the fracture strain for both the armchair and zigzag was ~25% and 18%, respectively which was noticeably different between the two orientations. However, the images of Figure 4 at a higher temperature suggest that the fracture strain range was 13-15% for both armchair and zigzag. This suggests that the fracture strain values were near each other at higher temperatures in comparison to lower temperatures. This difference can be attributed to increased thermal vibration of atoms at higher temperatures [72].

The stress-strain relations of both armchair and zigzag orientations for 10K and 300K are shown in Figure 5. Equation 5 was used to calculate the plane stress using virial stress data from the MD simulation. To validate our computational results, we compared our data (stress-strain) with existing literature work from [51] at high and low temperatures as shown in Supplementary Table 1. Our results were in good agreement with ref [51]

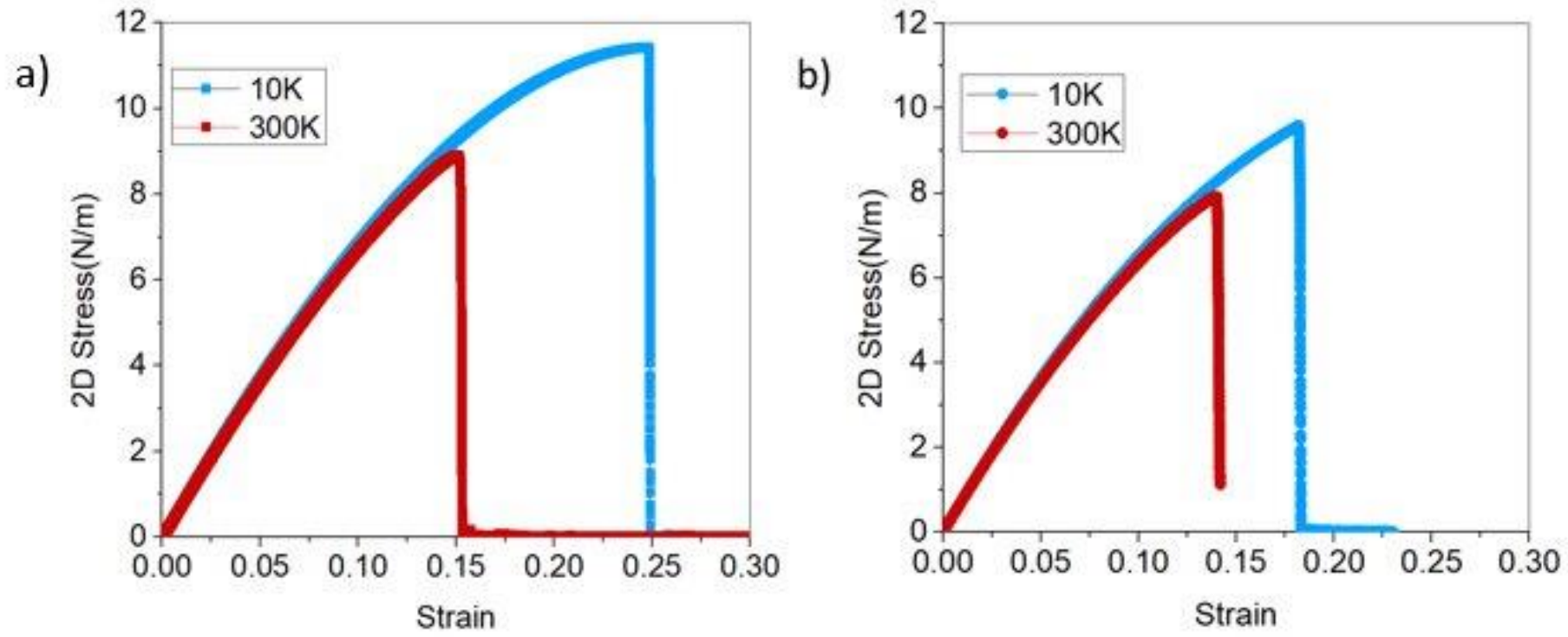

Figure 5: Stress-strain curve of a monolayer $MoTe_2$ under uniaxial tension along (a) armchair and (b) zigzag directions at 10K and 300K.

Furthermore, we investigated the effect of varying temperatures on the fracture behavior of monolayer $MoTe_2$ under uniaxial tension. In addition to our simulations at temperatures 10K and 300K, we also simulated the fracture behavior for 100K to 600K temperatures, with an interval of 100K. At higher temperature ranges, the effect of applied strain had a significant impact on the fracture strain of 2D sheets as shown in Figure 6.

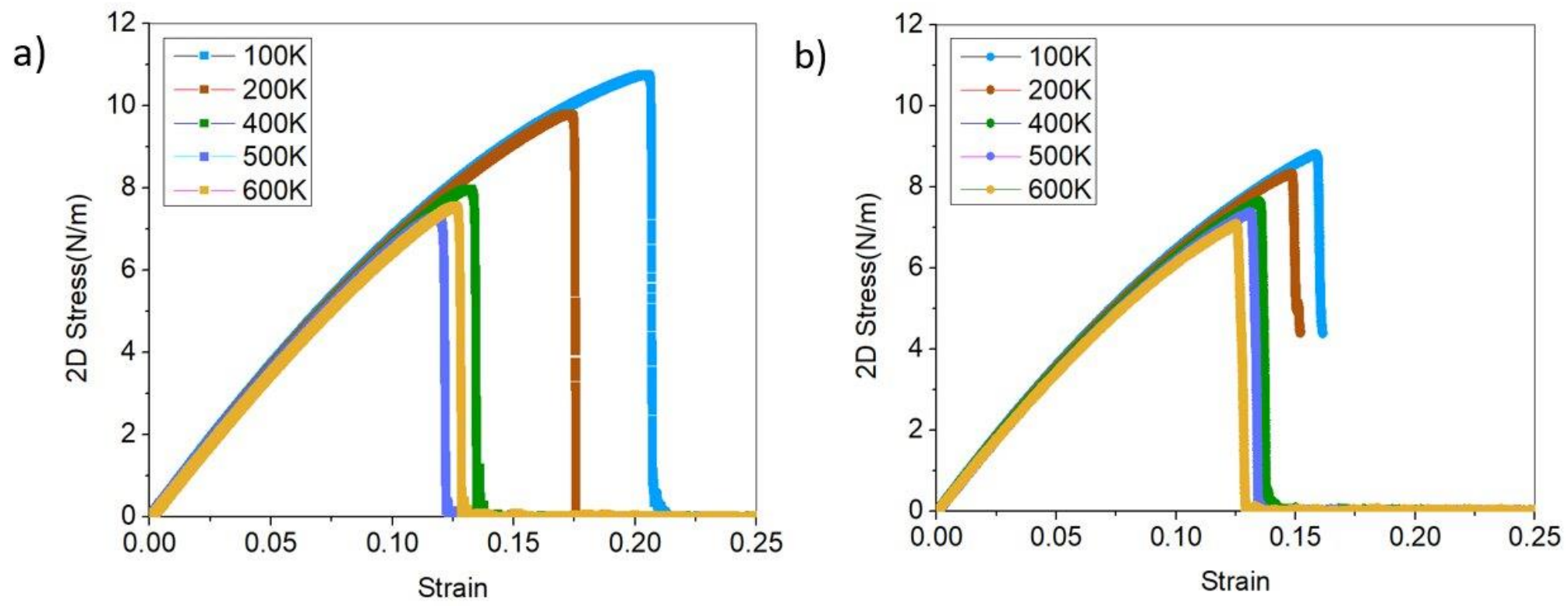

Figure 6: Stress-strain curve of a monolayer $MoTe_2$ under uniaxial tension along (a) armchair and (b) zigzag directions at different temperatures

Note that, even at elevated temperatures, the 2D sheet undergoes a brittle fracture without showing any sign of plastic deformation. Figure 6 also shows that the ultimate strain falls with increasing temperature due to the presence of high thermal vibrations [72]. The fracture strength of $MoTe_2$ was reduced by ~34% and ~26% in the armchair and zigzag direction, respectively with the increase in temperature. The fracture strain was reduced by ~49.5% and ~31% in the armchair and zigzag directions, respectively. Young's modulus was calculated by linear fitting of the stress-strain relation of the elastic region obtained by the virial stress equation. The fracture strength and Young's Modulus are significantly lower in the case of the zigzag direction in comparison to the armchair direction. The relation between the 3D Young's modulus, 3D fracture strength, fracture strain, and temperature are shown in Figure 7 (a-c).

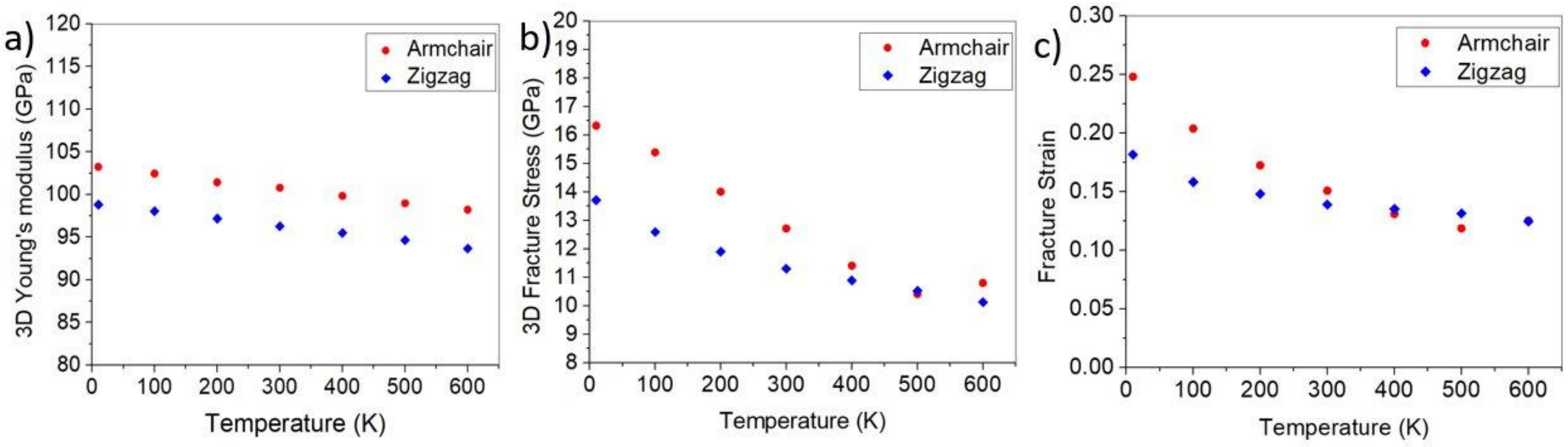

Figure 7: (a) Elastic modulus, (b) Fracture strength, and (c) Fracture strain of monolayer $MoTe_2$ at different temperatures.

To provide a comprehensive understanding of the research findings, we compared our obtained results with previously reported mechanical properties of monolayer TMDs. In Table 1, we reported the Young's Modulus and Fracture Strength of various TMDs with armchair edges at a temperature of 300K. The data recorded was obtained from MD simulations and experimental techniques. These results contribute to a comprehensive understanding of the mechanical properties of such materials, enhancing insight into their potential applications.

Table 1: The Young's Modulus and Fracture Strength of TMDs for an armchair edge at 300K temperature using experimental and MD simulations.

| 2D Materials | Thickness nm | Phase | Young's Modulus, Y (GPa) | Young's Modulus, Y N/m | Fracture Strength, $\sigma_{max}$ (GPa) | Computational Method | Potential Used |
|---|---|---|---|---|---|---|---|
| $MoS_2$ | 0.615 | Hexagonal | 158 | 97.17 | 12.5 | MD Simulation [73] | SW Potential |
| | 0.61 | Hexagonal | 155 | 94.55 | 11.5 | MD Simulation [74] | SW Potential |
| | 0.642 | Hexagonal | 159 | 102.1 | 15.03 | MD Simulation [70] | SW Potential |
| | 0.6-0.7<br>0.6-0.7 | Hexagonal<br>Trigonal | 135<br>107 | 87.75<br>69.55 | 12.00<br>8.0 | MD Simulation [53] | SW Potential |
| $WS_2$ | 10 (bulk) | Hexagonal | 107 | n/a | 19.16 | MD Simulation [75] | SW Potential |
| | 0.626 | Hexagonal | 182 | 113.9 | 19.46 | MD Simulation [76] | SW Potential |
| $WSe_2$ | n/a | n/a | 119 | n/a | 16.20 | MD Simulation [64] | SW Potential |
| | 0.67 | Hexagonal | 126 | 84.42 | 15.95 | MD Simulation [77] | Tersoff Potential |
| $MoSe_2$ | 0.6-0.7<br>0.6-0.7 | Hexagonal<br>Trigonal | 154<br>114 | 100.1<br>74.1 | 17<br>7 | MD Simulation [53] | SW Potential |
| $MoTe_2$ | 0.6-0.7<br>0.6-0.7 | Hexagonal<br>Trigonal | 121<br>107 | 78.65<br>69.55 | 18<br>8 | MD Simulation [53] | SW Potential |
| | 0.7 | Hexagonal | 110 | 77 | 14.5 | MD Simulation [39] | SW Potential |
| | 0.7 | Hexagonal | 102 | 71.4 | 12.85 | This Work (MD Simulation) | SW Potential |

## 4. Conclusion

The fracture behavior of a monolayer of molybdenum ditelluride ($MoTe_2$) at a wide range of temperatures was investigated using molecular dynamics simulations. The fracture experiment was simulated under varying temperatures for both armchair and zigzag directions. Our conclusions are:

(a) Temperature plays a significant role in the fracture mechanics of $MoTe_2$. The stress-strain plot depicts that their ultimate tensile strength decreased with increasing temperatures. However, we have also observed that at elevated temperatures the atomic orientation does not affect the fracture behavior significantly. This fracture behavior will play a key role in the utilization of these 2D materials for high-temperature applications.

(b) The visual results of the simulation reveal the difference in crack formation for armchair and zigzag directions. Uniaxial stress in armchair directions projects a clean and fine crack on the sheet. However, several distorted cracks were initiated by the same strain applied in a zigzag direction.

(c) It is found that Young's Modulus, fracture strength and fracture strain are anisotropic along armchair and zigzag orientations.

(d) Future scope to work on mechanical properties of multi-stacked, bulk $MoTe_2$ and defective $MoTe_2$.

**Author contributions:** ASM R. H and A. F. A. H modelled the experiments in the simulator and conducted the experiments. ASM R. H and A.F.A. H wrote the manuscript. M.T.F helped to run the simulations. M.R.K, S.S, M.T.F and M.A.I helped to review the manuscript. M.R.K, S.S, Y.A, M.T.F and M.A.I supervised the work. All authors discussed the results and reviewed the final manuscript.

**Competing interests:** The authors declare that they have no competing interests. Data and materials availability: All data needed to evaluate the conclusions in the paper are present in the paper. Additional data related to this paper may be requested from the authors.

**Acknowledgement:** The authors would like to acknowledge Prof. Jin-Wu Jiang from Shanghai Institute of Applied Mathematics and Mechanics, Shanghai University, China; Rafsan Al Shafatul Islam Subad from Department of Mechanical Engineering, University of Maryland, USA; and Dr. Arash Mobaraki from Bilkent University, Turkey for their valuable guidance and support.